\documentclass[12pt,fleqn]{tsp}

\techscience{vol.}{no.}{2024}{0}
\license{\includegraphics[width=16cm,height=1.5cm]{license}}
\thehead{Journal not specified}%
\newtheoremstyle{note}
{3pt}
{3pt}
{\normalfont}
{}
{\bfseries}
{.}
{.5em}
{}
\graphicspath{{./image/}}
\runningtitle{Manuscript Preparation for TSP} 
\usepackage{cite}
\usepackage{amsmath,amssymb,amsfonts}
\usepackage{graphicx}
\usepackage{authblk}
\usepackage{hyperref}
\usepackage{url}          
\usepackage{xurl}
\usepackage{caption}
\usepackage{lineno}
\usepackage{textcomp}
\usepackage{xcolor}
\usepackage{float}
\usepackage{amsmath}
\usepackage{algpseudocode}  
\usepackage[ruled,linesnumbered]{algorithm2e}

\title{Optimizing System Latency for Blockchain-Encrypted Edge Computing in Internet of Vehicles}

\author{Cui Zhang\thanks{School of Internet of Things Engineering, Wuxi Institute of Technology, wuxi, 214121, China}, Maoxin Ji\thanks{School of Internet of Things Engineering, Jiangnan University, wuxi, 214122, China}, Qiong Wu{$\textsuperscript{\textcolor{cyan}{\tiny 2}}$$^{\bm ,}$$^\ast{}$}, Pingyi Fan\thanks{Department of Electronic Engineering, Beijing National Research Center for Information Science and Technology, Tsinghua University, Beijing 100084, China} and Qiang Fan\thanks{Qualcomm, San Jose, CA, 95110, USA
 \vskip 0.15cm
 {$^\ast{}$Corresponding Author: Qiong Wu. Email: qiongwu@jiangnan.edu.cn}
 \vskip 0.15cm
 {Received: Month Day, Year; Accepted: Month Day, Year}}}

\begin{document}
\maketitle
\begin{mdframed}[backgroundcolor=gray!8,linewidth=0pt,nobreak=true,innerleftmargin=0.2cm,innerrightmargin=0.2cm,leftmargin=-2cm,rightmargin=-2cm]
\begin{abstract}
As Internet of Vehicles (IoV) technology continues to advance, edge computing has become an important tool for assisting vehicles in handling complex tasks. However, the process of offloading tasks to edge servers may expose vehicles to malicious external attacks, resulting in information loss or even tampering, thereby creating serious security vulnerabilities. Blockchain technology can maintain a shared ledger among servers. In the Raft consensus mechanism, as long as more than half of the nodes remain operational, the system will not collapse, effectively maintaining the system's robustness and security. To protect vehicle information, we propose a security framework that integrates the Raft consensus mechanism from blockchain technology with edge computing. To address the additional latency introduced by blockchain, we derived a theoretical formula for system delay and proposed a convex optimization solution to minimize the system latency, ensuring that the system meets the requirements for low latency and high reliability. Simulation results demonstrate that the optimized data extraction rate significantly reduces system delay, with relatively stable variations in latency. Moreover, the proposed optimization solution based on this model can provide valuable insights for enhancing security and efficiency in future network environments, such as 5G and next-generation smart city systems.
\end{abstract}

\end{mdframed}

\vspace{9pt}
\noindent\textbf{Nomenclature}
\vspace{3pt}

\noindent\begin{tabular}{@{}ll}
IoV & Internet of Vehicles\\
V2X & Vehicle-to-Everything \\
AI & Artificial Intelligence \\
DAGIoV & Directed Acyclic Graph-based Internet of Vehicles \\
BIoV & Blockchain Internet of Vehicles \\
V2V & Vehicle-to-Vehicle \\
RSU & Roadside Unit \\
BS & Base Station \\ 
BVIB & Blockchain-Enabled Variational Information Bottleneck \\
SNR & Signal-to-Noise Ratio 
\end{tabular}

\section{Introduction}
\label{intro}

With the continuous development of vehicle intelligence and connectivity, the amount of data that vehicles need to process has increased sharply. This data includes sensor data, map information, user interaction data, and more, which place higher demands on the computational capabilities of vehicles \cite{zhuang2019sdn, wang2024value}. However, due to limitations imposed by onboard hardware conditions and power consumption constraints, a single vehicle's computational power is insufficient, and the volume of data that can be processed is limited. Typically, it cannot meet the computational service requirements needed for task demands \cite{luo2023edgecooper, 10536013}. To address this issue, reliance on the Internet of Vehicles (IoV) network is essential.

IoV technology can effectively support vehicle-to-everything (V2X) communication \cite{10697115}. With the advancement of mobile edge computing technology, when a vehicle's own computing power is inadequate to handle complex local tasks, significant computational components can be offloaded to roadside units and base stations through the IoV, thereby reducing the local computational burden of the vehicle \cite{10663259}. As large artificial intelligence (AI) models develop and are progressively applied to IoV, the computational capacity of vehicles has become unprecedentedly strained. Many cutting-edge architectures have proposed using edge computing to provide AI services to vehicles. However, using AI to process local tasks involves the transmission of a large amount of private data. Many attackers might intercept or tamper with information during data transmission, severely affecting vehicular tasks and potentially causing serious security hazards \cite{8976295, 10643168}.

In vehicular networks, blockchain technology can establish secure data links, addressing the issue of data transmission security in vehicular networks \cite{cheng2019space}. When deploying blockchain technology, servers are typically set up as nodes within the blockchain network. These servers generally possess sufficient computational resources and storage capacity to handle large amounts of data and computational tasks \cite{zhang2022blockchain}. Servers collect various data from vehicles and package them into blocks. Each block contains a certain number of data records and is linked to another block, forming a reliable chain-like data structure that makes data in any block difficult to alter \cite{wu2022characterizing}. Therefore, integrating blockchain technology with edge computing can, to some extent, ensure the security of vehicular networks \cite{WANG2025103031}.

Traditional blockchain technologies, such as the Paxos consensus algorithm, are known for being difficult to understand and accurately implement. In 2011, Diego Ongaro and John Ousterhout proposed a simple and easy-to-understand distributed consensus algorithm named the Raft consensus mechanism \cite{184040}, which is also not inferior to the Paxos algorithm in terms of security. Building on this, we aim to combine the Raft consensus mechanism with edge computing technologies to create a transmission framework that is easy to understand and effectively enhances system security, alleviating the local computing power limitations of vehicles while ensuring secure information transmission.

However, the introduction of blockchain technology introduces additional time delays to vehicular network communications under ideal conditions. The Raft consensus mechanism requires periodic leader elections within the network to access and establish blocks. Followers need to send the received information to the leader and wait for the leader to aggregate the information and broadcast the generated block to them. These processes inevitably introduce additional communication delays including the time spent on task scheduling \cite{WANG2025107576}. To meet the ultra-low latency and ultra-high reliability requirements of vehicular communications, it is necessary to study the components of system delay and control them within an appropriate range through optimization methods.
\setcounter{footnote}{0}
In this paper, we outline a framework that combines blockchain technology and edge computing within IoV and optimize the system's latency based on this framework\footnote{The source code can be found at https://github.com/qiongwu86/BVIB-for-Data-Extraction-Based-on-Mutual-Information-in-the-IoV}. The specific contributions are as follows:

\begin{itemize}
	\item We enhance the security of network information transmission processes by integrating the Raft consensus mechanism within the IoV-based edge computing framework.
	
	\item Based on our proposed framework, we derive expressions for time delays in the process and design an online optimization algorithm based on convex optimization to ensure minimum system latency.
\end{itemize}

The rest of this paper is organized as follows: Section II summarizes the related research work, laying the foundation for this study. Section III describes the system model, including the environment and data extraction 
process. Section IV elaborates on the edge computing framework integrated with the Raft consensus mechanism. Section V derives the system delay and optimizes the latency using convex optimization methods.Section VI presents the simulation results. Section VII concludes the paper.

\section{Related Work}
Certain studies have explored the application of blockchain solutions to alleviate the computational burden in the vehicle-to-vehicle network. Wei et al. designed a confusion strategy to offload encrypted computation to other devices, reducing the computational burden in the vehicle-to-vehicle(V2V) network \cite{wei2021secure}. Nguyen et al. utilized smart contracts paired with double Q-networks to decrease the computational load on mobile devices within vehicle-to-vehicle networks, leveraging blockchain \cite{nguyen2021secure}. Lan et al. integrated drones into these networks, employing blockchain to ease computational demands \cite{10131971}.
In article \cite{8964493}, a Directed Acyclic Graph-based Internet of Vehicles (DAGIoV) framework is proposed for the Blockchain Internet of Vehicles(BIoV) system. This framework utilizes the Tangle data structure, treating each node as a miner, and achieves consensus among nodes using the Internet of Things Application (IOTA) consensus mechanism. Additionally, a game-theoretic approach is employed to determine the optimal service providers, thereby enabling low-cost communication.
Article \cite{9146574} focuses on providing security for IoV during spectrum sensing and information transmission based on cognitive radio networks. Blockchain technology is leveraged to maintain and track stored information within the network. This approach addresses the challenges of spectrum sensing and data sharing among available vehicles during mobility.
In article \cite{8998397}, a hybrid blockchain mechanism is introduced, which optimizes participating nodes through asynchronous federated learning. This optimization accelerates the learning rate, alleviates transmission loads, and addresses privacy concerns of vehicles. The proposed mechanism facilitates efficient data sharing within the vehicular network environment.
Article \cite{Qureshi2022ABE} combines blockchain technology to propose a privacy protection and authentication mechanism for network conditions in vehicular networks. Vehicles are treated as nodes, and Hyperledger Fabric is utilized to establish the blockchain. The framework provides multiple decentralized trusted authorities, avoiding single points of failure common in traditional networks. Additionally, it enables the tracking of malicious nodes through feature tracing, ensuring the system's security.
In article \cite{Li2022AnIC}, a novel distributed consensus algorithm suitable for vehicular networks is presented. Furthermore, a blockchain-based V2V cooperation mechanism is designed to prevent congestion. This approach enhances the efficiency and reliability of data transmission within the vehicular network.

Reference \cite{10308013} explores the application of blockchain technology in cloud, fog, and edge computing services. It suggested that blockchain can provide a trustless environment for data sharing and storage in edge computing, where edge devices serve as nodes to enable decentralized control. Reference \cite{8605952} proposed a VANET security architecture based on blockchain and mobile edge computing. In this architecture, blockchain is employed at the perception and service layers to ensure the security of data transmission between vehicles. The edge computing layer between the perception and service layers provides computational resources and edge cloud services, offering guidance for the integration of blockchain and edge computing in network frameworks. Reference \cite{9680338} investigated how to improve the efficiency of blockchain networks based on a combined blockchain and edge computing framework. By constructing a discrete-time Markov chain (DTMC) model to evaluate the performance of blockchain-based networks, it identifies key factors influencing network efficiency and determines strategies to optimize it. Reference \cite{10075287} optimized energy consumption and computational overhead in an IoT framework that combines blockchain and vehicular edge computing through the allocation of computational resources. By optimizing the profit between task rewards and weight costs, it can reduce the dimensionality of the action space and enhance system performance.

These studies mentioned above considered the integration of blockchain into vehicular networks to secure information transmission within the network. However, they do not address the time delay from the perspective of information transmission. This paper further investigates the relationship between vehicle data extraction rates and time delays, and optimizes the total system delay through convex optimization methods.

\begin{figure}[t]
	\centering
	\includegraphics[width=0.6\columnwidth, trim=0.5cm 0.5cm 0.5cm 0.5cm, clip]{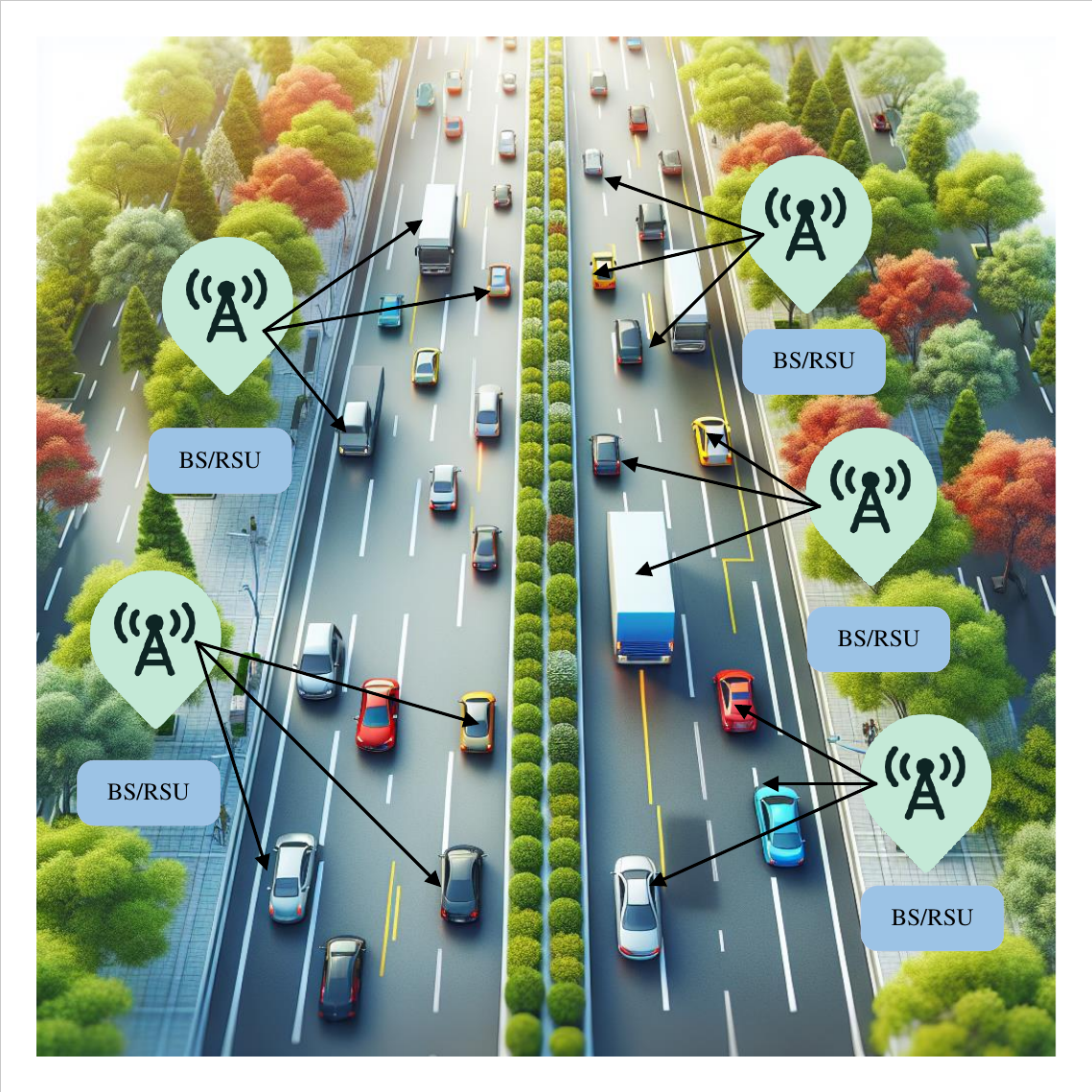}
	\caption{Environment.}
	\label{fig1}
\end{figure}

\section{System Model}
In this section, we construct a network architecture that integrates numerous vehicles with limited computational resources with servers possessing abundant computational capabilities, such as Road Side Units (RSUs) and base stations (BSs). Additionally, we provide a concise overview of the data extraction process performed by vehicles within the environment and the channel models involved in data transmission.
\subsection{Environment Description}
This study considers a scenario involving straight roadways, where RSUs or BSs are strategically deployed on both sides to assist with vehicular computations. Vehicles are required to offload certain computational tasks to these edge-deployed devices to alleviate the burden on their local processing capabilities. To ensure the security of information and prevent the tampering of data transmitted by vehicles, we establish a blockchain environment between the edge devices (i.e., BSs and RSUs) and implement the Raft consensus mechanism to defend against external attacks.

Specifically, each vehicle communicates with a single edge device. These edge devices can dynamically assume one of three roles: follower, candidate, or leader. The leader is responsible for receiving requests from vehicles and creating new log entries, which are uniquely appended to the leader's log. Followers are tasked with recording and storing log information received from the leader and responding to the leader's requests as necessary. Candidates can solicit votes from other nodes during the election process to establish their eligibility as the new leader; they serve as intermediate roles transitioning from followers to leaders 21. A visual representation of the system model is presented in Fig.~\ref{fig1}.

\subsection{Data Extraction Process}
The process of vehicle data extraction can be considered as a random arrival process over time. A Poisson process \( R \) is used to model the vehicle data extraction process. This modeling approach simulates the task demands of vehicles in vehicular networks. The Poisson point process has the following properties:

\begin{itemize}
	\item For a Poisson point process \( A([a,b)) \), the number of data points arriving within the time interval \([a,b)\) is determined by a Poisson random variable, with a mean of \( \lambda([a,b)) \).
	\item For different time intervals \([a_1, b_1), [a_2, b_2), \dots, [a_k, b_k)\), the corresponding Poisson point processes \( A([a_1, b_1)), A([a_2, b_2)), \dots, A([a_k, b_k)) \), with the number of data points \( N([a_1, b_1)), N([a_2, b_2)), \dots, N([a_k, b_k)) \), are independent random variables. These variables do not influence each other and each has its own expected value \( \lambda([a_1, b_1)), \lambda([a_2, b_2)), \dots, \lambda([a_k, b_k)) \).
\end{itemize}

For a point process on a non-negative time interval, let \( \{A(i)\}_{i \geq 1} \) represent the arrival times of the points. These points, when arranged in ascending order of their arrival times, are given by \( A_1 \leq A_2 \leq A_3 \leq \cdots \). The arrival times of these points are not necessarily uniformly distributed, and their order is part of the random process. The time intervals between adjacent points represent the inter-arrival times, which can be described as \( \mathcal{T}_i = A_i - A_{i-1} \), where \( A_0 = 0 \).

For a homogeneous Poisson process, the data arrival rate \( \Lambda \) is constant. Assuming the length of the time interval is \( t \), the average number of arrivals can be calculated using the following expression:
\begin{equation}
	\lambda(t) = \Lambda t.
	\label{eqa1.1}
\end{equation}

The corresponding inter-arrival times follow an independent and identically distributed exponential distribution with mean \( 1/\Lambda \), and the probability distribution is given by:
\begin{equation}
	\mathbb{P}(\mathcal{T}_i \leq t) = 1 - e^{-\Lambda t}.
	\label{eqa1}
\end{equation}

For a non-homogeneous Poisson process, the data arrival rate \( \Lambda \) is a time-varying function. The distribution of the first arrival time can be derived based on the arrival rate function \( \Lambda(t) \) of the non-homogeneous Poisson process, as follows:
\begin{equation}
	\mathbb{P}(\mathcal{T}_1 \leq t_1) = 1 - e^{-\int_0^{t_1} e^{-\Lambda(x)} dx}.
	\label{eqa2}
\end{equation}

Similarly, the probability distribution of the second arrival time can be described as:
\begin{equation}
	\mathbb{P}(\mathcal{T}_2 \leq t_2 \mid \mathcal{T}_1 \leq t_1) = 1 - e^{-\int_{t_1}^{t_2} e^{-\Lambda(x)} dx}.
	\label{eqa3}
\end{equation}

For \( i = 3, 4, \dots \), the distributions follow a similar form. The average number of arrivals in the time interval \( [0, t) \) for the non-homogeneous Poisson process can be calculated as:
\begin{equation}
	\lambda(t) = \lambda([0,t)) = \int_0^t \Lambda(t) \, dt,
	\label{eqa5}
\end{equation}
where \( \lambda(t) \) represents the cumulative arrival rate for all events from time 0 to time \( t \). This integral represents the total number of events over the interval \( [0, t] \), reflecting the accumulated arrival rate function \( \Lambda(t) \).

Moreover, in a non-homogeneous Poisson process, the distribution of each arrival point is independent, and the distribution of arrival points within a specific time interval follows a particular probability function. For any point \( a \in [0,t) \), the following formula holds:
\begin{equation}
	\mathbb{P}(A_i \leq a) = \frac{\lambda(a)}{\lambda(t)},
	\label{eqa4}
\end{equation}
where \( \lambda(a) \) is the cumulative arrival rate in the interval \( [0, a] \), and \( \lambda(t) \) is the cumulative arrival rate in the interval \( [0, t] \).

\begin{table}[H]
	\centering
	\caption{\textbf{Parameter Definitions Used in This Study.}}
	\newcolumntype{C}{>{\centering\arraybackslash}X}
	\begin{tabularx}{\textwidth}{C C}
		\toprule
		\textbf{Parameter} & \textbf{Symbol} \\
		\midrule
		Number of communication stations & $N$ \\
		Maximum vehicles served per station & $M$ \\
		Rate of data extraction by vehicles & $\lambda$ \\
		Velocity of vehicles & $v$ \\
		Carrier frequency of signal & $f_c$ \\
		Amount of Doppler shift & $f_d$ \\
		Probability of channel degradation & $P_p$ \\
		Probability of optimal channel condition & $P_i$ \\
		Probability of channel drop & $p_d$ \\
		Probability of collision occurrence & $p_c$ \\
		Delay in extracting data & $T_{ex}$ \\
		Delay in encoding vehicle data & $T_{ec}$ \\
		Transmission delay for data & $T_{m}$ \\
		Delay in data arrival & $T_{ar}$ \\
		Delay in decoding data at node & $T_{dc}$ \\
		Transmission delay from node to leader & $T_{f}$ \\
		Broadcasting delay & $T_{p}$ \\
		Election decision delay & $T_{ele}$ \\
		\bottomrule
	\end{tabularx}
	\label{tab:1}
\end{table}

\section{Secure Framework Based on the Raft Consensus Mechanism}
In our previous work, we proposed a Blockchain-Enabled Variational Information Bottleneck (BVIB) technique\cite{10468615}. In this section, we extend this framework by combining the Raft consensus mechanism within the blockchain with edge computing in vehicular networks. This integrated framework is applicable to various edge computing tasks and ensures the security of information.

In conventional edge computing technologies, vehicles offload computationally intensive tasks from their local systems to edge servers. However, when subjected to attacks by malicious nodes, this process can lead to information loss or even data tampering, significantly impacting local vehicular tasks. To address this, we incorporate the Raft consensus mechanism from blockchain technology to protect the information uploaded by vehicles. Additionally, tasks are distributed across different edge servers to expedite computation and provide timely feedback to the vehicles. Within our framework, the most critical components are the leader election process and the log maintenance procedure inherent to the Raft consensus mechanism.

\subsection{Election Process}
In the Raft consensus mechanism, all nodes are assigned one of three roles: leader, candidate, or follower. Initially, all nodes start as followers, each with a randomly assigned election timeout to minimize the probability of election conflicts. If a follower does not receive a heartbeat from the leader within its election timeout, it automatically transitions to a candidate and initiates a new voting round. The candidate sends vote requests to all nodes, which then agree to or reject the vote based on predefined rules. A candidate that secures a majority of votes becomes the new leader.

The leader is responsible for periodically sending heartbeat messages to all nodes within its term to assert its leadership. If a follower detects a heartbeat from a leader with a higher term, it reverts to a follower state. In scenarios where multiple candidates initiate elections simultaneously, vote splits may occur. In such cases, the election times out, and all candidates re-enter the election process until one obtains a majority of votes. If the leader fails, followers trigger a new election to ensure that there is always a leader within the cluster.

\subsection{Log Maintenance}
\begin{figure}[h]
	\centering
	\includegraphics[width=\columnwidth, trim=0.5cm 0.5cm 0.5cm 0.5cm, clip]{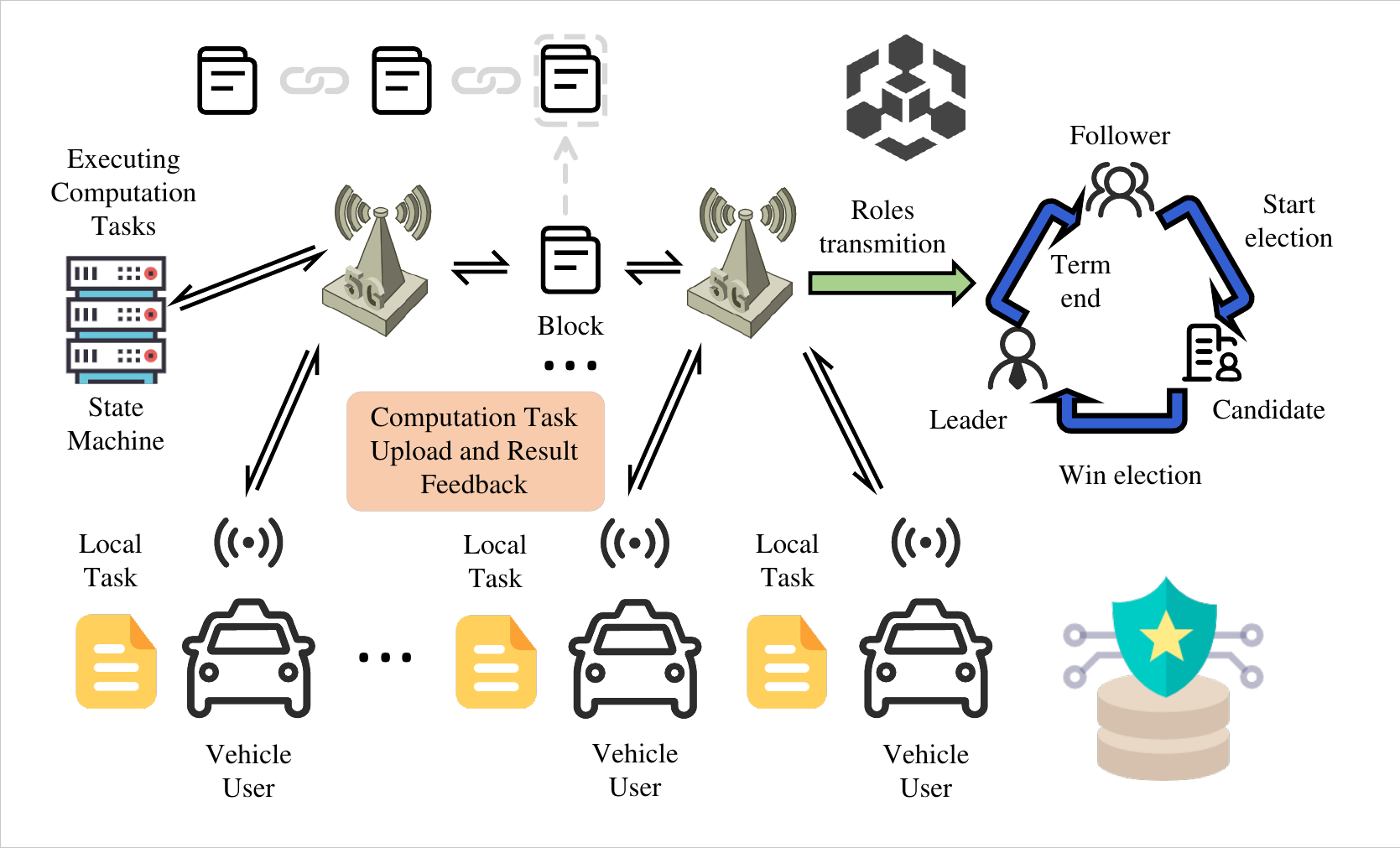}
	\caption{Secure Framework Based on the Raft Consensus Mechanism.}
	\label{fig2}
\end{figure}
In our framework, all edge-deployed servers (e.g., base stations and RSUs) function as nodes, while vehicles act as clients that send requests. When a vehicle performs edge computing, the computational task is uploaded as a request to the leader. The leader maps this request to a log entry and replicates it to all follower nodes. Upon receiving the log, each follower first verifies the log's validity (based on the leader's term number). After successful replication, followers send a confirmation back to the leader. Once a majority of followers have confirmed, the log entry is marked as committed, and the leader applies it to the state machine. Each node then executes the corresponding computational task based on the vehicle's location within its range, updates the task status upon completion, and returns the result to the vehicle, storing the information in the log.

Since logs are replicated across all nodes, the Raft consensus mechanism exhibits strong robustness against malicious attacks. If a leader is compromised and fails to send heartbeat messages promptly, a new election process is quickly initiated to elect a new leader. When some nodes are attacked and unable to respond in a timely manner, they can swiftly recover by retrieving the latest logs from the leader once normalcy is restored, thereby continuing operations seamlessly. Furthermore, when the leader sends logs and heartbeat messages, it verifies whether the followers' logs are matched. Only followers with matching logs receive the leader's log information, effectively preventing malicious nodes from intercepting or tampering with the data.

Fig.~\ref{fig2} illustrates the combined process of the Raft consensus mechanism and edge computing. Initially, a vehicle sends a computation task to a neighboring base station (RSU), which forwards it to the base station (RSU) acting as the leader at that time. The leader adds the collected task information to the log and replicates it to all nodes. After all nodes confirm, they return confirmation messages. Once more than half of the nodes have confirmed, the leader commits the log and assigns the task to the state machines of the other nodes. After each node completes the computation task, it adds the relevant information to the log and returns the computation results to the requesting vehicle.

The process of the entire framework is described in Algorithm 1. First, the leader node initializes a term and synchronizes it with all follower nodes. Whenever a vehicle V submits a computation task request, the request is forwarded to the leader node through the base station. The leader node generates a log entry and broadcasts it to all follower nodes. Each follower node validates the log entry, and if the term matches, it stores the log entry and sends an acknowledgment to the leader node. Once the leader node receives acknowledgments from more than half of the follower nodes, it marks the log entry as committed and executes the task in the state machine. Finally, the execution node performs the computation task based on the task's region and, upon completion, returns the results to the vehicle and updates the task status. The entire process ensures consistency and reliability in task scheduling and execution through the Raft protocol.

\begin{algorithm}[t]
	\caption{Task Scheduling and Log Maintenance with Raft Consensus Mechanism}
	\setstretch{0.75}
	Randomly initialize the term for the leader node $\texttt{term} = 1$;\\
	Initialize the term for all follower nodes to match the leader's term;\\
	\For{each computation task \( t_i \in T \)}{
		Vehicle \( V \) sends a computation task request \( t_i \) to the nearest communication base station;\\
		The base station forwards the request to the leader node \( n_{\text{leader}} \);\\
		The leader node \( n_{\text{leader}} \) performs the following operations: \\
		\Indp
		Generate the log entry \texttt{log\_entry} = (\( t_i \), \texttt{term});\\
		Store \texttt{log\_entry} in the local log;\\
		Send \texttt{log\_entry} to all follower nodes \( n_{\text{follower}} \in N_{\text{followers}} \);\\
		\Indm
		\For{each follower node \( n_{\text{follower}} \in N_{\text{followers}} \)}{
			Upon receiving \texttt{log\_entry}, the follower validates it: \\
			\If{\( \texttt{log\_entry.term} = n_{\text{follower}}.\texttt{term} \)}{
				The follower stores \texttt{log\_entry} in the local log;\\
				The follower sends a confirmation message to the leader;\\
			}
		}
		\If{The leader receives confirmations from more than half of the follower nodes} {
			The leader marks the \texttt{log\_entry} as \texttt{committed};\\
			The leader applies the task request to the state machine and executes the task;\\
		}
	}
	
	\For{each node \( n \in N \)}{
		\If{node \( n \) is an execution node (i.e., the node matches the vehicle's task region)}{
			Execute the computation task \( t_i \);\\
			Upon completion, update the task status to \texttt{completed};\\
			Send the computation result back to vehicle \( V \);\\
		}
	}
\end{algorithm}

\section{Delay Calculation and Optimization}
In this section, building upon our previously proposed edge computing framework integrated with blockchain technology, we analyze the system's time delays and employ convex optimization methods to minimize these delays.
\subsection{Channel Model}
Communication delay arises from data transmission between vehicles and servers \cite{1207121}. In this study, there are two distinct processes: uploading local vehicle tasks and receiving feedback from edge computing results. Within our blockchain-based framework, evaluating communication delays (both uploading and receiving feedback) requires an accurate analysis of the cellular network's channel dynamics, such as those related to V2V and V2X LTE \cite{deng2023reconfigurable,yue2024hybrid}. We conduct an approximate analysis by considering channel state transitions, as detailed in \cite{TMC_Pokhrel}, encompassing the core concepts.

Assume that the transmission time of a vehicle data block is equal to the duration of a single discrete time slot in the environment. Due to factors such as fading, the channel state in the environment fluctuates, and we model it as a binary state change. When the channel state is above a certain threshold, the channel is in an optimal state, and we assume that all transmissions are successful. During each transmission, the channel has a probability \( p_i \) of maintaining its current state and a probability \( 1 - p_i \) of transitioning to a worse state. When the channel state deteriorates and the channel is in a poor state, we assume that all transmissions will fail. Similarly, during each transmission, the channel has a probability \( p_p \) of maintaining its current state and a probability \( 1 - p_p \) of transitioning to an ideal channel state. Therefore, each transmission can be described by a fixed success probability.

Assume that the maximum channel capacity for each frame at the transmission link layer is \( \theta \). Due to the high-speed movement of vehicles, Doppler shift can be calculated by \( f_d = f_c v / c \), where \( v \) is the vehicle speed, \( f_c \) is the carrier frequency, and \( c = 3 \times 10^8 \text{ m/s} \). It is assumed that based on a specific modulation and coding scheme, the signal transmitted by the vehicle has some resistance to channel fading. Let the fading margin be \( F \). When the signal-to-noise ratio (SNR) at the receiver exceeds the threshold \( E[SNR] / F \), the channel is in the ideal state; otherwise, the channel is in the adverse state. Therefore, the average failure probability for frame transmission can be expressed as: \(\bar{p}_e = 1 - e^{-1/F}.\) To build a more detailed mathematical model, define the parameter \( \eta = \sqrt{2 / (F(1 - \rho^2))} \), where \( \rho \) is the Gaussian correlation coefficient, representing the correlation of the channel fading amplitude at Doppler frequency \( f_d \). The Gaussian correlation coefficient \( \rho \) reflects the degree of fading correlation of the signal at frequency \( f_d \). The closer \( \rho \) is to 1, the stronger the correlation of signal fading. \( 1 / \theta \) represents the frame transmission time in the ideal state. The first kind of zero-order Bessel function is denoted as \( J_0 \). Therefore, the channel state transition probabilities can be expressed as:

\begin{equation}
	\left\{
	\begin{aligned}
		&P_p = 1 - \frac{\mathbb{Q}(\eta,\rho\eta)-\mathbb{Q}(\eta,\rho\eta)}{e^{1/F}-1}, \\
		&P_i = 1 - \frac{1-\bar{p}_e(2-P_p)}{1-\bar{p}_e}.\\		
	\end{aligned}
	\right.
	\label{pppi}
\end{equation}

In \cite{TCOM_Pokhrel}, the channel discard probability can be expressed as:
\begin{equation}
	p_d=v_L\frac{1-p_i}{2-p_p-p_i}+\frac{l_L}{1+\frac{1-p_i}{1-p_p}},
	\label{pd}
\end{equation}
where \( v_L \) and \( l_L \) represent the probabilities of transmission failure in adverse and ideal channels, respectively.

\subsection{Delay computation}
First, vehicles need to extract data, which incurs an extraction delay $T_{ex}$. According to the property of homogeneous Poisson processes \eqref{eqa1}), the expectation of $T_{ex}$ is,
\begin{equation}
	E[T_{ex}]=\frac{1}{\lambda}.
	\label{tex}
\end{equation}

After the vehicle extracts the data, assuming the vehicle needs to perform simple processing on the data, consider the encoding delay $T_{ec}$. The delay when the vehicle sends the data is: 
\begin{equation}
	T_m = T_{ex}+T_{ec}.
	\label{tm1}
\end{equation}

Assume that each base station communicates with $M$ vehicles, and data packet collisions may occur during data transmission by the vehicles. In the LTE CAT M1 specification, a transmission interval of 2 to 3 time slots is typically set to avoid collisions. Suppose a 3-time-slot interval is adopted, i.e., $\tau_c = 3 \tau_t$, where $\tau_t$ is the time slot interval. The collision probability can be expressed as:
\begin{equation}
	\begin{aligned}
		&p_c = 1-\prod^{M} Pr((T_{m_1}-T_{m_2})>\tau_c), \\
		&\forall m_1,m_2 \in [1,M] \quad\&\quad m_1 \neq m_2
		\label{pc1}
	\end{aligned}	
\end{equation}
where $T_{m_1}$ and $T_{m_2}$ represent the times when any two vehicles send data. Also, due to \eqref{tm1}, $T_{m_1}$ and $T_{m_2}$ are Poisson process delays plus a constant, thus still constituting a Poisson process.

Therefore, we have,
\begin{equation} 
	p_c = 1-e^{-\lambda M(M-1) \tau_c/2},
	\label{pc2}
\end{equation}

We assume that after a collision occurs, vehicles re-extract the data. Therefore, the updated delay for vehicle data transmission is,
\begin{equation}
	T_m = \frac{T_{ex}+T_{ec}}{1-p_c}.
	\label{tm2}
\end{equation}

Similarly, considering the probability of transmission failure in the channel (Eq. \ref{pd}), the delay for data to arrive at the server is,
\begin{equation}
	T_{ar} = \frac{T_m}{1-p_d} = \frac{T_{ex}+T_{ec}}{(1-p_c)(1-p_d)}.
	\label{tar}
\end{equation}

Then, the server decodes the data, resulting in a decoding delay $T_{dc}$, which is a constant. Followers upload decoded variables $\hat{Y}$ to the leader, giving a delay of $T_f$, also a constant. The leader generates and broadcasts the block to all followers, resulting in a delay of $T_p$, also a constant. Finally, the leader's election time $T_{ele}$ similarly affects the overall network delay.

Combining the above with \eqref{tar}, the overall network delay is given by,

\begin{equation}
	\begin{aligned}
		\mathbb{T} &= T_{ar}+T_{dc}+T_f+T_p \\
		&= \frac{T_{ex}+T_{ec}}{(1-p_c)(1-p_d)} +T_{dc}+T_f+T_p+T_{ele}.
		\label{tn}
	\end{aligned}
\end{equation}

Combining \eqref{tex} and \eqref{pc2}, we have,

\begin{equation}
	E[\mathbb{T}] = \frac{1/\lambda+T_{ec}}{1-p_d}e^{\lambda M(M-1) \tau_c/2}+T_{si}+E[T_{ele}].
	\label{tt1}
\end{equation}
where $T_{si} = T_{dc} + T_f + T_p$, which is for simplification purposes. $E[T_{ele}]$ represents the expected value of $T_{ele}$.

If there is no attack and each term is not interrupted, then the average election time overhead for one epoch is simply the expected value of $T_{ele}$, denoted as $E[T_{ele}]$.
\begin{equation}
	E[T_{ele}] = \frac{E[\mathbb{T}]}{T_{term}}\tau_{ele},
	\label{tele1}
\end{equation}
$\tau_{ele}$ is the time it takes to elect a leader. Assuming constant control information intervals, carrier frequencies, and other conditions, without loss of generality, we have $\tau_{ele}=\tau_b \log N$, where $N$ represents the server count and $\tau_b$ is a constant representing the election time.

For an attack with strength $a$ ($0 \leq a < \frac{N}{2}$), it means that out of $N$ servers, $a$ servers have been attacked and their computing power has been paralyzed. In our architecture, as long as the paralyzed servers are not the leader, the impact on the system is relatively small. If the leader is paralyzed, its control information stops, and an election will immediately begin. Assuming each server has an equal probability of being attacked, the probability of the leader being attacked is $\frac{a}{N}$. This results in additional election time overhead, so \eqref{tele1} is modified to be:
\begin{equation}
	E[T_{ele}] = \frac{E[\mathbb{T}]}{T_{term}}\tau_{ele}(1+\frac{a}{N}),
	\label{tele2}
\end{equation}

Combining \eqref{tele2}, \eqref{tt1} should be modified as follows,
\begin{equation}
	E[\mathbb{T}] = \frac{1/\lambda+T_{ec}}{1-p_d}e^{\lambda M(M-1) \tau_c/2}+T_{si}+\frac{E[\mathbb{T}]}{T_{term}}T_{ele}(1+\frac{a}{N}),
	\label{tt2}
\end{equation}

Combining equations \eqref{tele2} and \eqref{tt1}, we can revise as follows,
\begin{equation}
	E[\mathbb{T}] = \frac{NT_{term}}{NT_{term}-(N+a)T_{ele}}(\frac{1/\lambda+T_{ec}}{1-p_d}e^{\lambda M(M-1) \tau_c/2}+T_{si}),
	\label{tt3}
\end{equation}

\subsection{Dynamic optimization algorithm}

\begin{algorithm}[t]
	\SetKwInOut{Input}{Input}\SetKwInOut{Output}{Output}
	\BlankLine
	\While{vehicles work}{
		\If{$\lambda \neq \lambda^*$}{
			$\lambda  \leftarrow \lambda ^*$ using \eqref{lamdapeak2}\;
		}
		
		\If{$N \rightarrow N^{new}$}{
			Jump step2\;
		}
		\If{$M \rightarrow M^{new}$}{
			Jump step2\;
		}
	}
	\caption{Real-time data extra algorithm}
	\label{alg:rda}
\end{algorithm}

The goal for the vehicles is to find the most suitable $\lambda$ to minimize $E[\mathbb{T}]$. Observing \eqref{tt3}, we find it to be a convex optimization function with respect to $\lambda$. We can differentiate it with respect to $\lambda$ to find the extreme points. Since \eqref{tt3} is quite complex, we can initially substitute $A=\frac{NT_{term}}{NT_{term}-(N+a)T_{ele}}$ and $B=M(M-1) \tau_c/2$ for these two coefficients.
\begin{equation}
	E[\mathbb{T}] = \frac{A}{1-p_d}[({T_{ec}}+\frac{1}{\lambda})e^{B\lambda}+T_{si}],
	\label{tt4}
\end{equation}

To take its derivative,
\begin{equation}
	E'[\mathbb{T}]=\frac{Ae^{B\lambda}}{(1-p_d)\lambda^2}(BT_{ec}\lambda^2+B\lambda-1),
	\label{ttd}
\end{equation}

Observing that $\frac{-AE^{-B\lambda}}{(1-p_d)\lambda^2}$ is always non-zero, $E'[\mathbb{T}]$ can only be zero when $BT_{ec}\lambda^2+B\lambda+1$ equals zero, for $E[\mathbb{T}]$ to attain an extreme value. According to Vieta's theorem,
\begin{equation}
	\lambda^*=\frac{-B\pm\sqrt{B^2+4BT_{ec}}}{2BT_{ec}},
	\label{lamdapeak1}
\end{equation}

Since $\lambda > 0$, \eqref{lamdapeak1} has only one valid solution,
\begin{equation}
	\lambda^*=\frac{-B+\sqrt{B^2+4BT_{ec}}}{2BT_{ec}},
	\label{lamdapeak2}
\end{equation}

Observing that within the interval $(0,\lambda^*)$, $E'[\mathbb{T}]$ is less than 0, hence $E[\mathbb{T}]$ decreases; and within the interval $(\lambda^*,+\infty)$, $E'[\mathbb{T}]$ is greater than 0, thus $E[\mathbb{T}]$ increases. Therefore, at $\lambda=\lambda^*$, $E[\mathbb{T}]$ reaches its minimum.

When a vehicle finds that its data extraction rate doesn't match $\lambda^*$, it adjusts its extraction rate to $\lambda^*$.

\begin{table}[t]
	\centering
	\caption{\textbf{Summary of Simulation Parameters.}}
	\newcolumntype{C}{>{\centering\arraybackslash}X}
	\begin{tabularx}{\textwidth}{C C}
		\toprule
		\textbf{Parameter Name} & \textbf{Symbol} \\
		\midrule
		Leader Term Duration & $T_{\text{term}} = 2000$ \\
		Election Time & $\tau_{\text{ele}} = 150$ \\
		Number BS & $N = 10$ \\
		Number of Attacked Nodes & $a = 2$ \\
		Vehicle Encoding Delay & $T_{\text{ec}} = 10$ \\
		Collision Timeout & $\tau_{\text{c}} = 0.1$ \\
		Drop Probability & $P_{\text{d}} = 0.02$ \\
		Node Decoding Delay & $T_{\text{dc}} = 10$ \\
		Transmission Delay from Node to Leader & $T_{\text{f}} = 2$ \\
		Broadcast Delay & $T_{\text{p}} = 2$ \\
		\bottomrule
	\end{tabularx}
	\label{tab:sim_params}
\end{table}

\begin{figure}[t]
	\vspace{-15pt}
	\centerline{\includegraphics[width=0.8\textwidth]{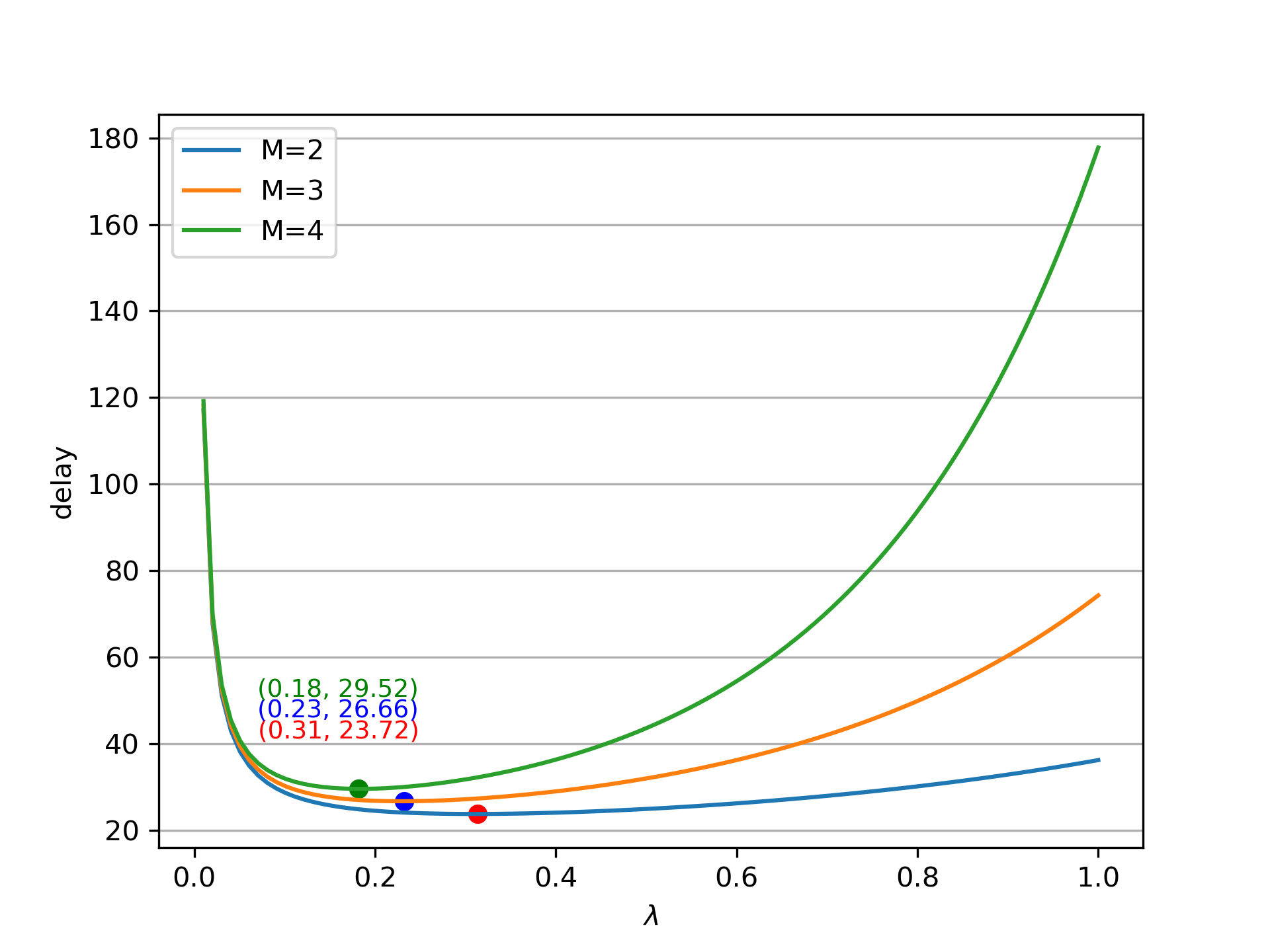}}
	\caption{latency with $\lambda$.}
	\label{fig4}
\end{figure}

\section{Performance Evaluation}
In this section, we employ Python3.8 for our simulation environment, utilizing communication network configurations that adhere to the specifications of 3GPP LTE Cat. M1 rules\footnote{see technical reports at https://www.3gpp.org/DynaReport.}. In this section, we use Python in a simulation environment, employing communication network configurations that comply with the 3GPP LTE Cat standards. To evaluate the performance of the proposed method, we measure the system's time delay by adjusting parameters such as the vehicle data extraction rate, the attack intensity in the environment, the number of vehicles, and the maximum number of connections per base station. Since being under attack may cause additional delays, the gains from system security are already reflected in the delay calculations.

Fig.~\ref{fig4} illustrates the relationship between system latency and the rate parameter $\lambda$. Depending on the different numbers of $M$, the optimal points of $\lambda$ vary. All three curves demonstrate that there exists an optimal point for system latency, which aligns with our theoretical derivations. In the case of $M=2$, the curve for $\lambda$ exhibits a rapid decrease followed by a rapid increase, with the optimal point at 0.31. Similarly, for $M=3$, the curve initially decreases rapidly then increases rapidly, with the optimal point at 0.23. In the case of $M=4$, the curve behaves similarly, with the optimal point at 0.18. When $\lambda$ is either too low or too high, the system latency increases. This is because when $\lambda$ is sufficiently small, the result in Equation \eqref{tt4} approaches infinity. Physically, when the data extraction rate of the vehicles is close to zero, the time required for the data extraction task increases without bound. When $\lambda$ approaches 1, as the number of vehicles increases, the probability of information collision between vehicles rises, leading to higher latency. Moreover, as the number of vehicles increases, the optimal point shifts to the left, and the optimal value increases. This is because the impact of information collision on latency becomes larger than the effect of data extraction. Therefore, it is necessary to ensure that the data extraction rate $\lambda$ is within a reasonable range.

\begin{figure}[t]
	\vspace{-10pt}		
	\centerline{\includegraphics[width=0.8\textwidth]{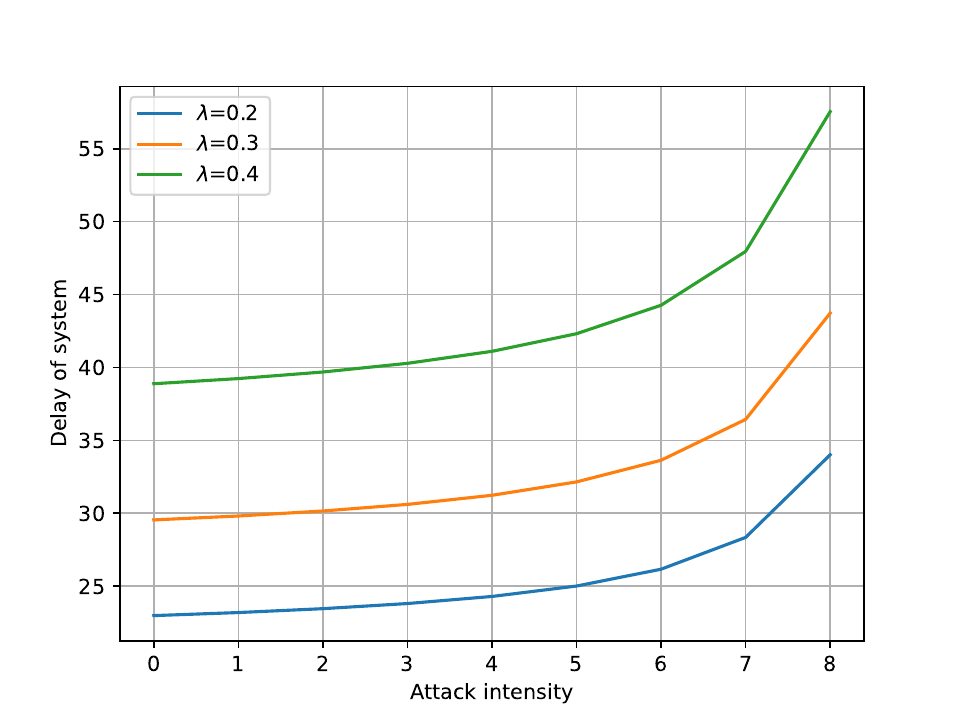}}
	\caption{Latency under attack.}
	\label{fig5}
\end{figure}
\begin{figure}[t]
	\vspace{-10pt}
	\centerline{\includegraphics[width=0.8\textwidth]{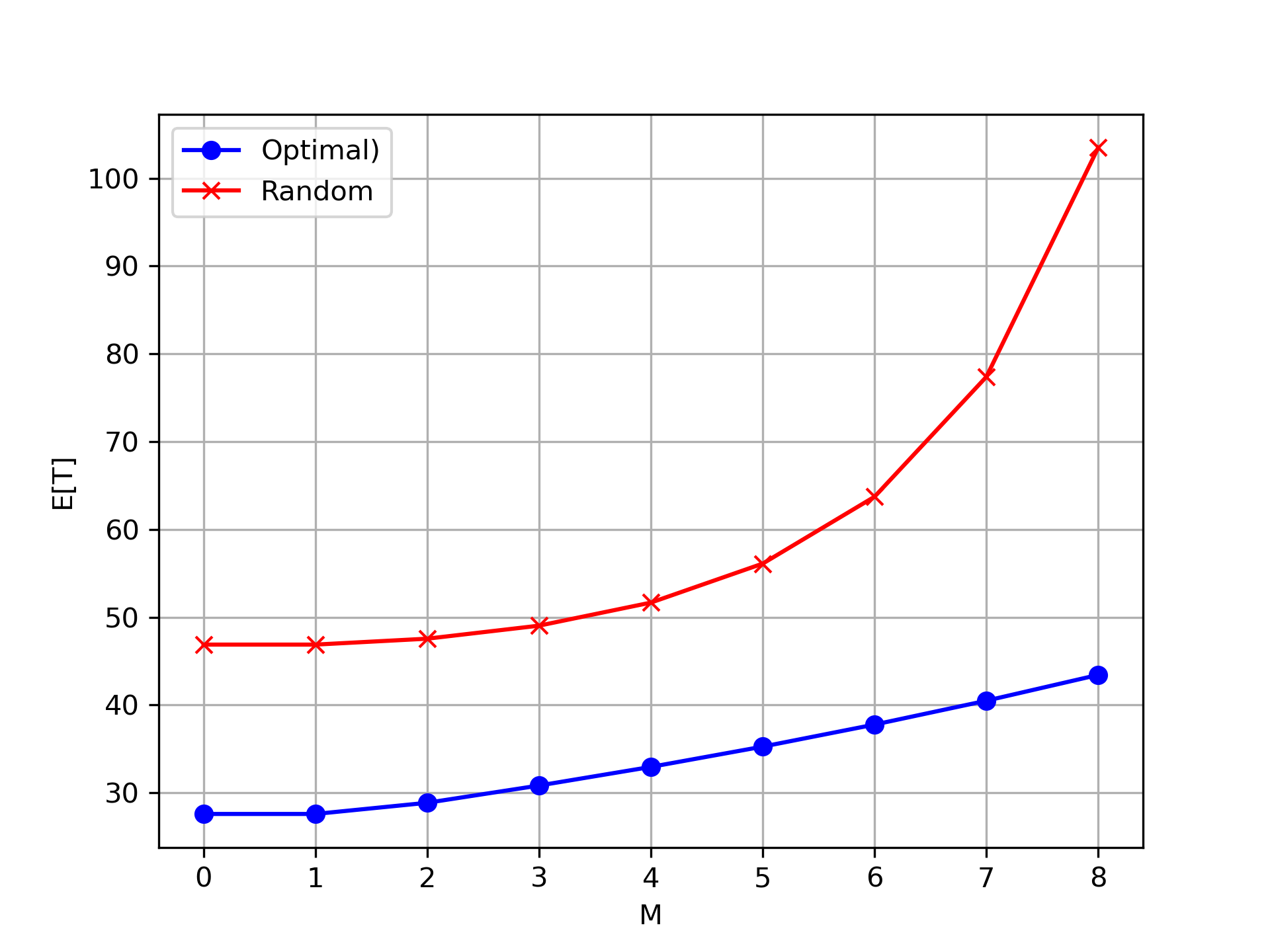}}
	\caption{Comparison of average delay between Random Algorithm and Optimal Algorithm with different number of vehicles.}
	\label{fig6}
\end{figure}
\begin{figure}[t]
	\vspace{-10pt}
	\centerline{\includegraphics[width=0.8\textwidth]{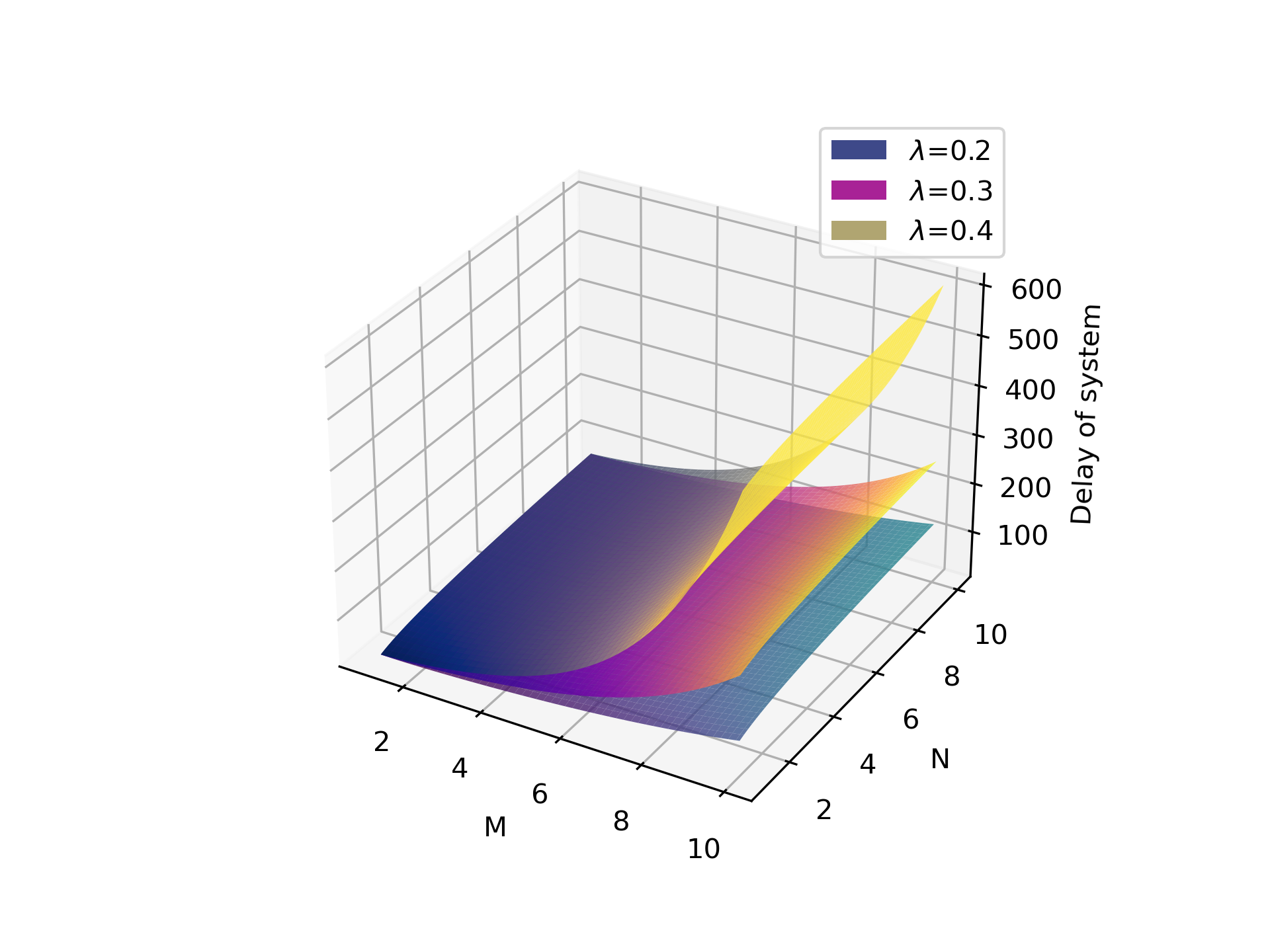}}
	\caption{Latency with $M$ and $N$.}
	\label{fig7}
\end{figure}

Fig.~\ref{fig5} illustrates the impact of attack intensity on system latency when $M=3$. The attack intensity is simulated by disabling servers, where an intensity of 1 represents disabling one server, and an intensity of 2 represents disabling two servers. The figure shows that as the attack intensity increases, the system latency also increases continuously. Before the attack intensity reaches 5, the increase in latency is not significant. However, after this point, the latency sharply rises, exhibiting $a$ trend similar to an exponential function. According to Equation \eqref{tt3}, the number of attacked nodes, denoted as $a$, primarily affects the probability of the leader node being paralyzed. A higher $a$ value results in a worse system state and larger latency. A comparison between the two curves reveals that the overall performance is much better when $\lambda$ is set to 0.2 compared to when it is set to 0.3 or 0.4.

Fig.~\ref{fig6} compares the system's average latency when using a random algorithm to determine $\lambda$ and when using the optimal algorithm proposed in this work, for different vehicle numbers $M$. The average value for the random algorithm is obtained by 1000 random experiments. As the number of vehicles increases, the system's average latency gradually rises. This is because the probability of information collision between vehicles increases, leading to higher latency. When using the random algorithm, the average information latency is significantly higher than with the proposed optimal algorithm, and the rate of increase in average latency is faster as the number of vehicles increases. Additionally, due to the random selection of $\lambda$, the system's average latency exhibits significant fluctuations. The average of 1000 random experiments reveals a long-term stable trend, but in real-time scenarios, its performance is much worse than that of the targeted optimization method.

Fig.~\ref{fig7} illustrates the simultaneous impact of $M$ and $N$ on system latency. The graph demonstrates that the system's latency rises in conjunction with increases in both $M$ and $N$. However, the rate of increase in system latency accelerates with increasing $M$, while it gradually slows down with increasing $N$. This is due to the influence of the system latency formula. When $M$ increases, the probability of information collision between vehicles rises significantly, whereas the $N$ value affects the election time, which follows a logarithmic model.  Similarly, the overall performance is much better when $\lambda$ is set to 0.2 compared to when it is set to 0.3 or 0.4.

\section{Conclusion}

This paper proposes an edge computing architecture that integrates blockchain technology and designs an online optimization algorithm based on the relationship between vehicle data extraction rate and system latency, achieving the optimal solution through convex optimization. During the computation, error-prone channels are considered to closely replicate real communication scenarios. Our conclusions are summarized as follows:

\begin{itemize}
	\item In our research, we propose an architecture that combines blockchain technology with edge computing, specifically designed to enhance the security and performance of the IoV environment. By incorporating the Raft consensus mechanism, we enhance the robustness and stability of the system when faced with channel errors and malicious attacks. To address potential security risks during the process of vehicles offloading tasks to edge servers, we utilize the immutable and distributed ledger characteristics of blockchain to ensure the system's security and the integrity of information.
	
	\item We establish an online optimization algorithm based on convex optimization to minimize the system's latency. Our algorithm effectively optimizes using precise delay expressions and vehicle data extraction rates, confirming its applicability and efficiency under various real communication environments. Furthermore, simulation results show that this method not only reduces system latency but also ensures efficient data transmission and processing capabilities, thereby meeting the stringent requirements of low latency and high reliability in the IoV.
\end{itemize}

In conclusion, our approach not only provides new perspectives and solutions theoretically but also offers practical guidance for real-world deployment, aiding the further development of edge computing and blockchain technology applications in the IoV. Our research holds significant implications for optimizing the security and effectiveness of IoV systems in the future.

\vspace{+6pt}
\noindent
\textbf{Acknowledgement:} We are grateful for the encouragement and support from our families and friends.

\vspace{+6pt}
\noindent
\textbf{Funding Statement:} This work was supported in part by the National Natural Science Foundation of China under Grant No. 61701197, in~part by the National Key Research and Development Program of China under Grant No. 2021YFA1000500(4), and in~part by the 111 project under Grant No.~B23008.

\vspace{+6pt}
\noindent
\textbf{Author Contributions:} Conceptualization, C.Z. and M.J.; methodology, C.Z., M.J. and Q.W.; software, C.Z. and M.J.; writing---original draft preparation, C.Z. and M.J.; writing---review and editing, P.F. and Q.F. All authors have read and agreed to the published version of the manuscript.

\vspace{+6pt}
\noindent
\textbf{Availability of Data and Materials:} Not applicable.

\vspace{+6pt}
\noindent
\textbf{Ethics Approval:} Not applicable.

\vspace{+6pt}
\noindent
\textbf{Conflicts of Interest:} The authors declare no conflicts of interest to report regarding the present study.

\bibliographystyle{IEEEtran}
\bibliography{tsp}

\end{document}